# Heliophysics Discovery Tools for the 21st Century: Data Science and Machine Learning Structures and Recommendations for 2020–2050

R. M. McGranaghan, B. Thompson, E. Camporeale, J. Bortnik, M. Bobra, G. Lapenta, S. Wing, B. Poduval, S. Lotz, S. Murray, M. Kirk, T. Y. Chen, H. M. Bain, P. Riley, B. Tremblay, M. Cheung, V. Delouille

Three main points: 1. Data Science (DS) will be increasingly important to heliophysics; 2. Methods of heliophysics science discovery will continually evolve, requiring the use of learning technologies [e.g., machine learning (ML)] that are applied rigorously and that are capable of supporting discovery; and 3. To grow with the pace of data, technology, and workforce changes, heliophysics requires a new approach to the representation of knowledge.

# 1   Introduction

We are at a crossroads in the study of Heliophysics. On one hand we operate in the same paradigm that has guided the field over the past couple of decades, ruled by the triumvirate of data, theory, and simulations. On the other hand, we are beginning to recognize that powerful new opportunities for scientific discovery are possible through increased data volume and sophisticated methods to explore these data. The emergence of the hyperconnected digital society and the massive quantities of data it generates has led to new analysis capabilities that scale well to the solar-terrestrial environment. Heliophysics is squarely positioned to benefit from the emerging field of data science [1].

In the context of this white paper we define *data science (DS)* as scalable architectural approaches, techniques, software and algorithms which alter the paradigm by which data are collected, managed, analyzed, and communicated. A key sub-field within DS is 'machine learning.' We follow [9] in defining *machine learning (ML)* to be computer algorithms that improve (or learn) from data and recognizing that they cover a wide spectrum—wide in complexity and application. Details of specific ML algorithms are being covered by companion white papers and an overview of Heliophysics applications is provided by [2].

This paper focuses on an important and poorly communicated side of DS and ML: the application of these methods as discovery tools. We provide two examples to trace this concept and to reveal what the Heliophysics community must prioritize in the 2020s to prepare for the following decades. Notwithstanding recent advances in the accuracy of ML models, the notion of *explainability* of models is crucial to Heliophysics and the wider artificial intelligence (AI) community. We offer a working definition with the primary goal to make this a central topic in the future of our field.

> **We make three central points to guide the Heliophysics community over the next three decades, specifically focusing on the rapidly evolving data and technology landscape and our utilization of it:**
>
> 1. **Data Science (DS) will be increasingly important to Heliophysics;**
>
> 2. **Methods of Heliophysics science discovery will continually evolve, requiring the use of learning technologies (e.g., machine learning (ML)) that are applied rigorously and that are capable of supporting discovery; and**
>
> 3. **To grow with the pace of data, technology, and workforce changes, Heliophysics requires a new approach to the representation of knowledge.**

## 2 How is this being done now? What are the guiding examples?

Our intention is not to provide a comprehensive set of examples, but to highlight those that illustrate the advent of DS and ML for discovery in the sciences and that frame explainability and recommendations for Heliophysics. A living resource of examples are maintained by the community at: https://tinyurl.com/Helio-DS-resources.

### 2.1 ML for scientific discovery: The NASA Frontier Development Laboratory

FDL The NASA Frontier Development Laboratory (FDL) applies ML technologies to science to push the frontiers of research and develop new tools to help solve some of the biggest challenges that humanity faces. Heliophysics has factored prominently in the applications addressed by FDL since its inception. FDL is a public-private partnership that removes computational resource restriction (e.g. through industry-provided cloud computational resources) and aligns teams of ML researchers and domain scientists to generate new scientific discovery. FDL, therefore, illustrates the need to converge sophisticated DS infrastructure, team diversity, and new approaches to knowledge integration.

### 2.2 ML rigor and community organization: The NASA Goddard Space Flight Center for HelioAnalytics (CforHA)

Advances in ML have been driven by industry. The availability of tools developed for corporate purposes has led to their adoption in the sciences often as 'black boxes.' Though these tools are indeed powerful for discovering patterns in data, in order to be tools of scientific discovery in Heliophysics, our community must develop an understanding of the principles of these models. This understanding is needed to build appropriately tuned ML models, to interrogate ML model performance, and to find ways to permit physics to inform ML models (e.g., [3, 4]). Growing sophistication of Heliophysics ML will depend on providing training to Heliophysicists and increasing interaction with the DS, computer science, and ML communities. Recently, the Center for HelioAnalytics at the NASA Goddard Space Flight Center has been established to create more rigorous ML for Heliophysics. The Center objectifies a community and information structure that permits DS tools and techniques to add depth and richness to the scope of scientific questions that can be asked and answered.

Despite the examples of relatively more structured environments of FDL and the CforHA, the majority of DS and ML in Heliophysics is still done by single researchers/groups. Though these groups may or may not be funded by NASA/NSF, they are too often more like 'spare time' projects. Finally, we note that equally important to the use of DS and ML for scientific discovery are the software and computational resources that they require. Open, adaptable, high capability platforms are inseparable from DS and ML progress.

## 3 Explainability

These examples build to an over-arching topic that will be increasingly important to Heliophysics and indeed all scientific fields: explainability. Explainability is the complement of methods and techniques that makes results of complex analyses (such as ML) understandable by humans (e.g., [5, 6]). We offer two foundational concepts of explainability that must be taken up by Heliophysics: 1) knowledge structuring and 2) uncertainty quantification. *Knowledge structuring* refers to the explicit interrelation of facts or knowledge about a particular topic. Improving the structuring of information is the key to integrating diverse concepts and data, developing traceable information models, and building discovery systems. The 'knowledge graph (KG)' [7] offers an intriguing concept to build explainability and we believe will be inextricable infrastructure for the future of Heliophysics. *Uncertainty quantification* (UQ) is the science of quantitative characterization and reduction of uncertainties. UQ is the means to achieve a paradigm shift for Heliophysics toward probabilistic approaches focused on the reliable assessment of uncertainties [2]. Both components are needed to: (i) move toward more generally intelligent systems (i.e., artificial intelligence, AI), and (ii) ensure space weather forecasts are accompanied by uncertainty estimates.

## 4 Recommendations[*]

---

[*] We acknowledge that challenges related to ML in Heliophysics are shared with other disciplines and this white paper

1. **Review and augment programs for data analysis.** The starting point to improve Heliophysics DS is to invest in the programs that support it. These most aptly fall into data analysis programs such as the NASA ROSES 'Heliophysics Data Environment Enhancements.' We recommend that these programs be reviewed to more deliberately identify and include techniques of ML and knowledge architecture.

2. **Produce incentive structures that value open source software and rigorous data science-for-science research in Heliophysics** The traditional incentive in our community is to publish research papers as the primary research artifact and to be the central metric of one's impact and contribution. The Heliophysics community can no longer flourish under these incentive structures alone. Though publications will remain important, their nature needs to change to become fluid, living, collaborative pieces (see, for a strong example, PubPub).

Additionally, non-traditional research artifacts, such as analysis software, metadata-rich data sets, and collaborative computational resources (such as Jupyter Notebooks) need to be incentivized. This will be an important shift in the Heliophysics research culture. In many respects, the progress we have made in DS and ML have relied on open-source scientific software. We must commit resources and time to supporting the open-source scientific software ecosystem.

3. **Increase support for Heliophysics data infrastructures.** The Space Physics Data Facility (SPDF) expands Heliophysics data usage and is inextricable from existing analyses, however it remains incomplete, lacking a formal Heliophysics ontology that links the data. We recommend investing resources to expand and enhance SPDF and its ability to link to other domains to support the transdisciplinary research that Heliophysics naturally requires.

4. **Build a Community of Practice (CoP) for Heliophysics that links traditionally disparate 'knowledge bases' and produces technologies capable to utilize the knowledge.** Knowledge structuring and the technologies to improve it is very much an open area of research. We must create a community that holds and advances this discussion in a sustained and transdisciplinary manner.

5. **Invest in education for new users of ML.** Educational resources (whether through courses or less formal structures) create new users of ML and raise awareness of the applications, both centrally important to a successful future of ML in Helio. We recommend continuation and expansion of programs like FDL, and investigation into additional programs to serve larger user groups similar to the AI for Earth System Science (AI4ESS) Summer School. Following [9], we imagine undergraduate and graduate Heliophysics science degrees incorporating courses in data science (see [8], about the need across the sciences for data rich courses). There will be a corresponding need for faculty in growth areas, and several funding programs have encouraged university investment through tenure track position sponsorship (see the NSF Faculty Development in the Space Sciences program) [9].

6. **Prioritize support for underrepresented members of our community**. It is imperative to acknowledge our field's role in structural racism, reexamine our biases and ways of operating, and participate in co-creating new strategies and support to actively recruit Black, Indigenous, and People of Color (BIPOC) to be members of the Heliophysics community through DS initiatives.

7. **Support transdisciplinary work.** We recommend new programs that are designed to fund research that fall into the gaps between traditional disciplines. This will require cross-agency coordination.

benefitted from that written by *Azari et al.*, [2020] for the Planetary Science Decadal Survey [9]. Several of our recommendations were influenced by that work.